\documentclass[aps,prl,twocolumn,showpacs,superscriptaddress,floatfix]{revtex4}

\usepackage{graphics}
\usepackage{epsfig}
\usepackage{epstopdf}
\usepackage{float}
\usepackage{upgreek}
\usepackage[T1]{fontenc}
\usepackage{mathptmx}
\usepackage{color}

\begin{document}


\title{High Pressure Superconducting transition in Dihydride BiH$_{2}$ with Bismuth Open-Channel Framework}

\affiliation{Beijing National Laboratory for Condensed Matter Physics and Institute of Physics, Chinese Academy of Sciences, Beijing 100190, China}  
\affiliation{School of Physical Sciences, University of Chinese Academy of Sciences, Beijing 100190, China}
\affiliation{Key Laboratory of Materials Physics, Ministry of Education, School of Physics, Zhengzhou University, Zhengzhou 450001, China}
\affiliation{Institute of Quantum Materials and Physics, Henan Academy of Sciences, Zhengzhou 450046, China}
\affiliation{Center for High Pressure Science and Technology Advanced Research, Beijing 100193, China}
\affiliation{Shanghai Synchrotron Radiation Facility, Shanghai Advanced Research Institute, Chinese Academy of Science, Shanghai 201204, China}

\author{Liang Ma$^{1,3,4, \#}$}
\author{Xin Yang$^{5, \#}$}
\author{Mei Li$^{5, \#}$}
\author{Pengfei Shan$^{1,2,5, \ast}$}
\author{Ziyi Liu$^{1,2}$}
\author{Jun Hou$^{1,2}$}
\author{Sheng Jiang$^{6}$}
\author{Lili Zhang$^{6}$} 
\author{Chuanlong Lin$^{5}$}
\author{Pengtao Yang$^{1,2}$}
\author{Bosen Wang$^{1,2}$}
\author{Jianping Sun$^{1,2}$}
\author{Yang Ding$^{5}$}
\author{Huiyang Gou$^{5}$}
\author{Haizhong Guo$^{3,4, \dagger}$}
\author{Jinguang Cheng$^{1,2, \ddagger}$}

\date{\today}

\begin{abstract}

Metal hydrides \textit{M}H$_{x}$ (x $\leq$ 2) with low hydrogen content are not expected to show high‑$T_\mathrm{c}$ superconductivity owing to the low hydrogen‑derived electronic density of states at Fermi level and the limited hydrogen contribution to electron-phonon coupling strength. In this work, we report on the successful synthesis of a novel bismuth dihydride superconductor, \textit{Cmcm}-BiH$_{2}$, at approximately 150 GPa, and the discovery of superconductivity with $T_\mathrm{c}$ about 62 K at 163 GPa, marking the first instance of superconductor among the \textit{M}H$_{2}$‑type metal dihydrides. \textit{Cmcm}-BiH$_{2}$ adopts a unique host-guest type structure, in which the Bi atoms via weak Bi-Bi covalent bonds form a three dimensional open-channel framework that encapsulates H$_{2}$-like molecules as guests, thereby broadening the structural diversity of hydrides under high pressures. The occurrence of superconductivity is evidenced by a sharp drop of resistivity to zero and the characteristic downward shift of $T_\mathrm{c}$ under applied magnetic fields. Notably, \textit{Cmcm}-BiH$_{2}$ remains stable down to at least 97 GPa during decompression, with the calculated lowest pressure for dynamic stability of 10 GPa. In-depth analysis reveals that the covalent bismuth open-channel structure forms metallic conduction channels, dominates the electronic states near the Fermi level, and contributes approximately  51\% of the total $\lambda$ in \textit{Cmcm}-BiH$_{2}$, distinguishing it from known high-pressure hydride superconductors. These findings highlight the critical role of non-hydrogen elements in producing superconductivity and open new avenues for the design and optimization of high-$T_\mathrm{c}$ hydride superconductors.

\end{abstract}

 
\maketitle

The pursuit of high-temperature superconductors has been a central goal in condensed matter physics since the discovery of superconductivity in solid mercury in 1911 \cite{A1}. According to the Eliashberg theory \cite{A2} and Allen-Dynes equation \cite{A3} for conventional BCS superconductors, the superconducting critical temperature ($T_\mathrm{c}$) is determined mainly by the electron-phonon coupling (EPC) strength and the frequency of the phonons involved in EPC. These factors are quantified by the EPC constant ($\lambda$) and the average frequency of the phonon spectrum weighted by the coupling strength ($\omega_\mathrm{log}$). As a result, light-element dominant compounds, especially hydrides, have long been considered as promising candidates for achieving high $T_\mathrm{c}$ due to inherently high phonon frequencies, which can not only elevate the Debye frequency and enhance the $\lambda$ but also boost intrinsic $\omega_\mathrm{log}$ values. Recent theory-driven experimental discoveries have indeed demonstrated near-room-temperature superconductivity in hydrides under high pressures, such as SH$_{3}$ \cite{A4}, LaH$_{10}$ \cite{A5, A6}, and CaH$_{6}$ \cite{A7, A8}. These hydride superconductors are featured by high hydrogen contents and three-dimensional hydrogen network, which facilities strong EPC, e.g., $\lambda$ $\sim$ 2 for LaH$_{10}$, via high-frequency H-H vibrations \cite{A9, A10}. Thus, the synergic enhancement of $\lambda$ and $\omega_\mathrm{log}$ based on the hydrogen sublattices is essential for the achievement of high $T_\mathrm{c}$ in these superhydrides superconductors.

Motivated by this concept, we have been considering whether a similar approach can be applied to design new superconductors with improved $T_\mathrm{c}$. It is well known that various bismuth (Bi) allotropes and compounds exhibits large $\lambda$ comparable to those superhydrides, e.g., the Pb-Bi alloy ($T_\mathrm{c}$ $\sim$ 9 K, $\lambda$ $\sim$ 2.3) \cite{A11}, amorphous Bi ($T_\mathrm{c}$ $\sim$ 6 K, $\lambda$ $\sim$ 2.5) \cite{A12}, and the Bi-\uppercase\expandafter{\romannumeral3} phase ($T_\mathrm{c}$ $\sim$ 7 K, $\lambda$ $\sim$ 2.75) \cite{A13}. The low-frequency phonons from heavy elements contribute to a large $\lambda$ value but result in a low $T_\mathrm{c}$ due to much reduced $\omega_\mathrm{log}$ in comparison to superhydrides. Then, an interesting question arises, \textit{i.e. whether the incorporation of hydrogen in bismuth can raise $T_\mathrm{c}$ via enhancing $\omega_\mathrm{log}$ while maintaining the large $\lambda$?} To this end, we recently initiated experimental explorations of bismuth hydrides in the Bi-H system under high pressures and obtained a novel molecular hydride superconductor, \textit{C}2\textit{/c}-BiH$_{4}$, with notably high $T_\mathrm{c}$ = 91 K at 170-190 GPa, setting a new benchmark for molecular hydride superconductors \cite{A14}. Unlike clathrate-type superhydrides that rely primarily on medium-to-high frequency hydrogen vibrations for EPC, the low-frequency translational vibrations of Bi atoms account for 38\% of the total $\lambda$ in the \textit{C}2\textit{/c}-BiH$_{4}$.

During the course of exploring the Bi-H systems, we observed another phase with a lower $T_\mathrm{c}$ $\sim$ 60 K in addition to \textit{C}2\textit{/c}-BiH$_{4}$. Here we report on the discovery and comprehensive characterizations of this novel bismuth dihydride, \textit{Cmcm}-BiH$_{2}$, synthesized at approximately 150 GPa and 2000 K. Interestingly, the structure of \textit{Cmcm}-BiH$_{2}$ is featured by a three-dimensional bismuth open-channel that encapsulates hydrogen molecules. Despite its low hydrogen content and the molecular nature of hydrogen in its structure, \textit{Cmcm}-BiH$_{2}$ was found to exhibit an unexpectedly high $T_\mathrm{c}$ of 62 K at 163 GPa, marking the first instance of superconductor in \textit{M}H$_{2}$‑type metal dihydrides. Our calculations revealed that the electronic density of states at the Fermi level,  \textit{N}($E_\mathrm{F})$, is predominantly contributed by the Bi 6\textit{p} orbitals and about 51\% of $\lambda$ arises from low-frequency vibrations of Bi atoms. This suggests that the observed high-$T_\mathrm{c}$ superconductivity in BiH$_{2}$ is mainly driven by the covalent network of Bi atoms, while the high-frequency vibrations from hydrogen molecules that enhance $\omega_\mathrm{log}$ significantly boost the superconducting properties. These findings not only underscore the distinctive nature of \textit{Cmcm}-BiH$_{2}$ compared to other high-pressure hydride superconductors but also offer valuable insights for the design and optimization of high-$T_\mathrm{c}$ hydride superconductors.

Based on our previous work on \textit{C}2\textit{/c}-BiH$_{4}$ \cite{A14}, we explored the Bi-H system at lower pressures from 150 to 170 GPa aided by laser heating to $\sim$ 2000 K and prepared a total of eight diamond anvil cells (DACs), using Bi foil and hydrogen sources, i.e., ammonia borane (BH$_{3}$NH$_{3}$), paraffin (C$_\mathrm{n}$H$_\mathrm{2n+2}$, n = 18-39) and pure molecular hydrogen as reactants. The samples were labeled as s1 and s8 for structure analysis and s2 to s7 for transport measurements. The XRD data was also collected for s5 and s6 after resistance measurements as detailed in Table S1 \cite{SI}. Further details on sample preparation, structural and transport measurements, and DFT calculations are provided in the Supplementary Materials (SM) \cite{SI}.

Figure 1(a, b) presents the synchrotron XRD patterns of s1 at selected pressures, where strong diffraction spots indicate a good crystallinity of the synthesized material. Comparison of the obtained XRD data with the Bragg positions of the cubic phase of bismuth clearly indicates that the bismuth metal almost fully reacted with the hydrogen released from the decomposition of BH$_{3}$NH$_{3}$. However, the XRD pattern cannot be described by any experimentally known or theoretically predicted bismuth hydrides \cite{A15}. Therefore, we employed the CALYPSO structure search method to identify the structure based on the obtained XRD pattern \cite{SI5, SI6, SI7}. Detailed procedures of the structure search are provided in the SM \cite{SI}. The results reveal that the diffraction peaks matched well with the BiH$_{2}$ phase with space group \textit{Cmcm}. As shown in Fig. 1(a), Rietveld refinement using the \textit{Cmcm}-BiH$_{2}$ structural model yields an excellent fit. The calculated equation of state (EOS) closely matches the experimental \textit{P-V} data shown in Fig. 1(c), further confirming that the structure is \textit{Cmcm}-BiH$_{2}$.

\begin{figure}[!t]
	\begin{center}
		\epsfxsize=8.5cm
		\epsffile{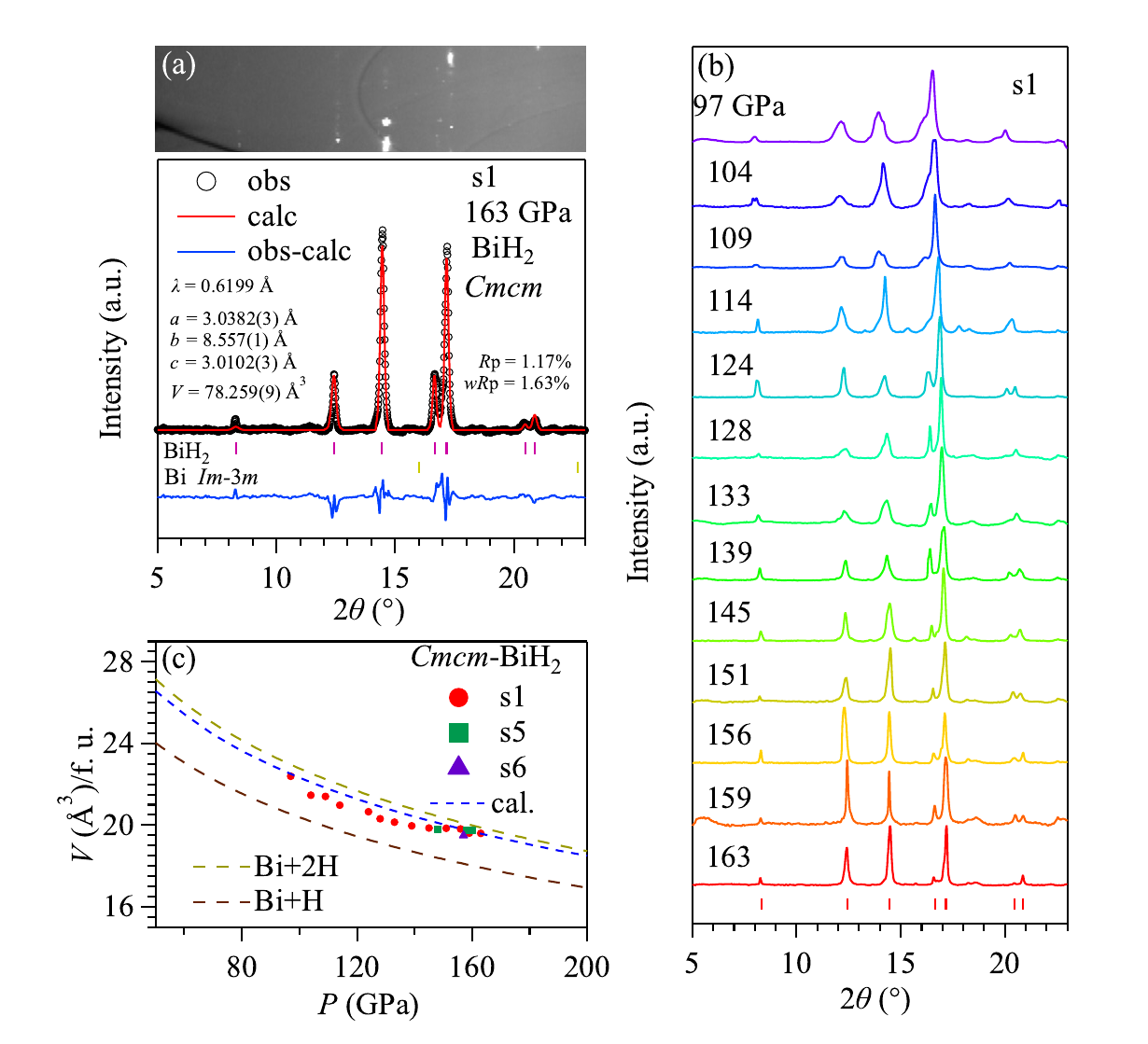}
	\end{center}
	\caption{(Color online) XRD patterns and the result of Rietveld refinement for BiH$_{2}$ of s1 at 163 GPa in the \textit{Cmcm} structure. Black circles: experimental data; red line: simulated XRD based on the structural model; purple and yellow vertical lines: Bragg diffraction positions of \textit{Cmcm}-BiH$_{2}$ and \textit{Im-}3\textit{m}-Bi, respectively; blue line: the difference between the simulated and the original XRD. The upper panel shows the XRD image of corresponding powder XRD pattern with the incident X-ray wavelength of 0.6199 {\AA}. (b) Representative integrated XRD patterns for BiH$_{2}$ of s1 upon decompression from 163 to 97 GPa. (c) Experimentally obtained volume per formula unit for BiH$_{2}$ plotted as a function of pressure. Red solid circles, green solid squares and purple triangle represent experimental data for BiH$_{2}$. Theoretical EOS of BiH$_{2}$ is plotted as blue dashed line. Brown and yellow dashed line represents ideal mixtures of Bi + H and Bi + 2H.}
	\label{fig1}
\end{figure}

Previous theoretical studies suggested that \textit{P}2$_{1}$\textit{/m}-BiH$_{2}$ has the lowest enthalpy for this stoichiometry between 117 and 200 GPa \cite{A15}. However, compared with experimental data, the XRD pattern of \textit{P}2$_{1}$\textit{/m}-BiH$_{2}$ shows significant discrepancies, as shown in Fig. S1 \cite{SI}. Then we carried out Gibbs free energy calculations under realistic high-pressure, high-temperature conditions to evaluate the relative stability of \textit{Cmcm}-BiH$_{2}$ and \textit{P}2$_{1}$\textit{/m}-BiH$_{2}$. The relative Gibbs free energy between \textit{Cmcm}-BiH$_{2}$ and \textit{P}2$_{1}$\textit{/m}-BiH$_{2}$ as a function of temperature and pressure are presented in the \textit{P-T} phase diagram shown in Fig. S2 \cite{SI}, which demonstrates that the formation of \textit{Cmcm}-BiH$_{2}$ is thermodynamically favored at elevated temperatures within the pressure range of 150-200 GPa. Detailed procedures of the calculations are provided in SM \cite{SI}. As shown in Fig. 1(b), the XRD patterns upon decompression down to 97 GPa do not show any peak splitting or appearance of new diffractions. This confirms that \textit{Cmcm}-BiH$_{2}$ synthesized at 163 GPa remains stable down to at least 97 GPa, below which the diamond anvils cracked during further depressurization process.

\begin{figure}[!t]
	\begin{center}
		\epsfxsize=8.5cm
		\epsffile{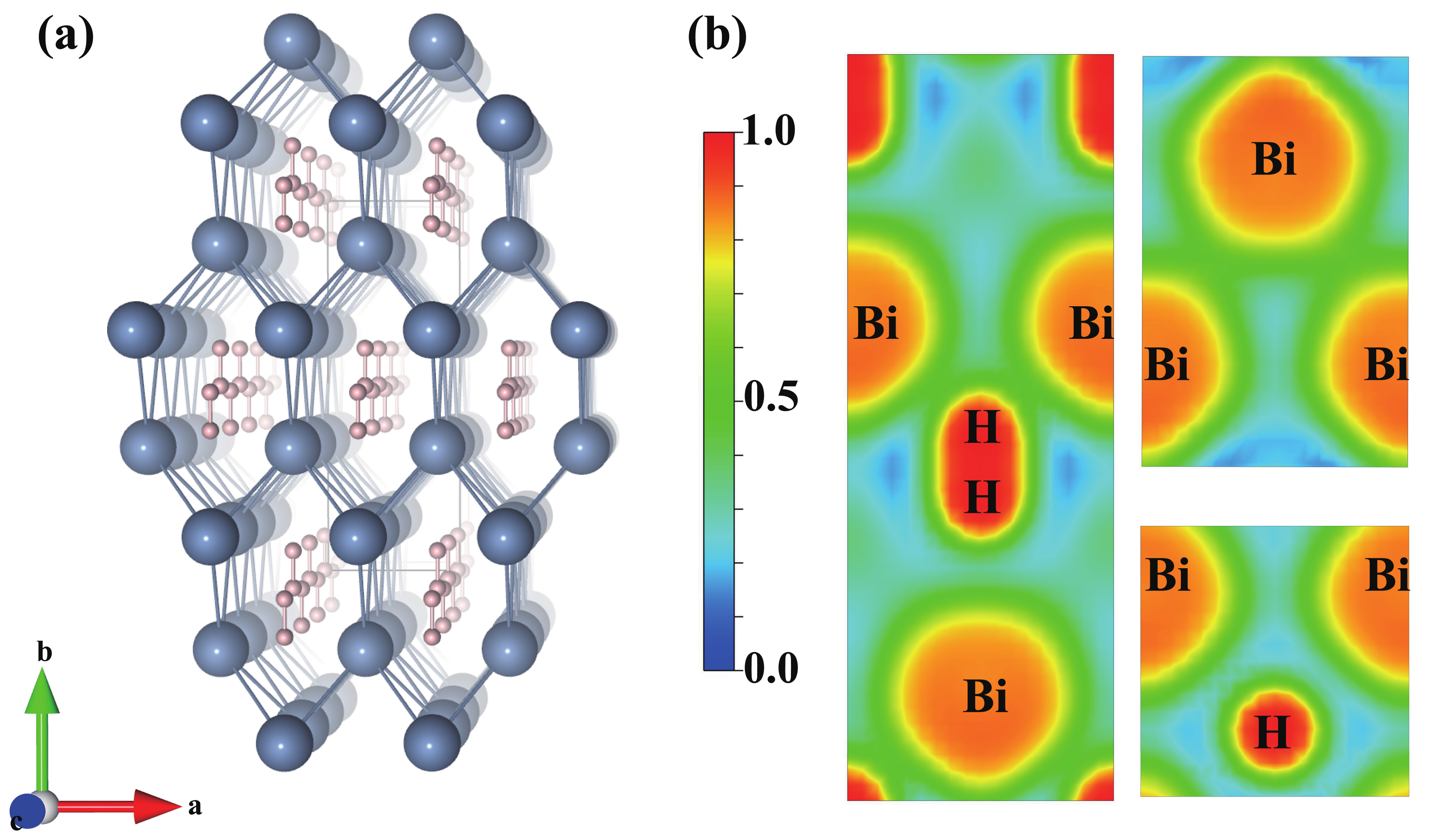}
	\end{center}
	\caption{(Color online) (a) Crystal structure of \textit{Cmcm}-BiH$_{2}$. Blue and pink spheres represent bismuth and hydrogen atoms, respectively. (b) The ELF projected on the plane (0 0 1) (left panel) for H-H contacts and the nearest Bi-H contacts, plane (0 2 1) for the nearest Bi-Bi contacts (up to the right panel), and plane (0 11 -1) for the sub-closest Bi-H contacts (down to the right panel) of BiH$_{2}$.}
	\label{fig2}
\end{figure}

Figure 2(a) shows the crystal structure of \textit{Cmcm}-BiH$_{2}$ for which the most prominent feature is the open channel framework of Bi atoms filled by H$_{2}$-like molecules, forming a host-guest type structure like the high-pressure phase of Bi-\uppercase\expandafter{\romannumeral3}. The calculated structural parameters are provided in Table S3 \cite{SI}. It is well known that the Bi-\uppercase\expandafter{\romannumeral3} phase adopts an intriguing host-guest structure \cite{A19}, characterized by a host lattice (\textit{I}4\textit{/mcm}) of covalently connected Bi atoms containing chains of guest Bi atoms (\textit{I}4\textit{/mmm}) with an incommensurate \textit{c}-axis relative to the host unit cell. It is noted that the Bi-Bi bond lengths in \textit{Cmcm}-BiH$_{2}$ (2.8-2.9 {\AA} at 150 GPa) are significantly shorter than those in the Bi-\uppercase\expandafter{\romannumeral3} phase (3.12-3.34 {\AA} at 6.8 GPa) \cite{A19}, which suggests the possible covalent bonding between Bi atoms in \textit{Cmcm}-BiH$_{2}$. For the H$_{2}$-like molecular units, the H-H bond length of $\sim$ 0.80 Å at 150 GPa is close to that in \textit{C}2\textit{/c}-BiH$_{4}$ \cite{A14}, and slightly elongated compared to that of a hydrogen molecule (0.74 {\AA}) at ambient pressure \cite{A20} but remains much shorter than those in monatomic hydrogen (0.98 {\AA} at 500 GPa) \cite{A21}. This further elaborates the molecular hydride nature of the structure, with the Bi atoms exerting a precompression effect on the H$_{2}$ molecules.

The bonding nature of \textit{Cmcm}-BiH$_{2}$ was first explored using electron localization function (ELF) and Bader charge analysis. At 150 GPa, the ELF value between the nearest-neighbor H-H atoms is approximately 0.97, confirming the presence of hydrogen molecules. In contrast, the ELF value between Bi-Bi atoms is relatively lower ($\sim$ 0.52), indicating the weak covalent bonding, which facilitates the formation of metallic conductive channels between Bi atoms, Fig. 2(b). Bader charge analysis reveals an effective charge transfer of approximately 0.32e per Bi atom from Bi to H, which slightly elongated the H-H intramolecular bond length. To gain further insights into bonding behavior, we employed crystal orbital Hamilton population (COHP) analysis to quantify interatomic interactions. The integrated COHP (ICOHP) and negative projected COHP (-pCOHP) data for intramolecular H-H, Bi-Bi, and Bi-H interactions at different pressures are presented in Table S4 and Fig. S3 \cite{SI}. The strong intramolecular H-H interaction observed through COHP analysis confirms the molecular hydride character of H$_{2}$ molecules. Additionally, the H-H bonds exhibit an antibonding character, as indicated by the negative -pCOHP at the Fermi level, supporting that electron transfer from Bi atoms to the H$_{2}$-like molecular units fills the antibonding orbitals. The ICOHP analysis reveals weak covalent interactions between Bi atoms, forming a distorted hexahedral prism with Bi atoms alternately arranged in triangular prismatic six-coordination units. Such configuration results in the formation of a three-dimensional bismuth open-channel structure. Unlike the clathrate hydride superconductors where a weak covalent hydrogen clathrate serves as the host and metal atoms act as guests, \textit{Cmcm}-BiH$_{2}$ exhibits a three-dimensional Bi atom framework with weak covalent Bi-Bi interactions as the host, encapsulating H$_{2}$-like molecules as guests. Such a bismuth host framework closely resembles the Bi-\uppercase\expandafter{\romannumeral3} phase, which was recently revealed to feature a significant large EPC, $\lambda$ $\sim$ 2.75 and non-Fermi-liquid behaviors prior to the superconducting transition at $T_\mathrm{c}$ $\sim$ 7 K \cite{A13}. In this regard, it is essential to conduct resistance measurements to examine the superconducting properties of \textit{Cmcm}-BiH$_{2}$.

Following the synthesis conditions for s1, we compressed s2 to 155 GPa at room temperature and subsequently laser-heated it to $\sim$2000 K, after which the pressure decreased to 152 GPa. We conducted independent experiments on s3, s4 and s5 under similar experimental conditions to check the reproducibility. The results are depicted in Fig. 3(a) and Fig. S4 \cite{SI}, respectively. Figure 3(a) presents the temperature-dependent resistance \textit{R}(\textit{T}) of s2 at 152 GPa, with the inset showing a photograph of the sample in the DAC used for resistance measurements. The \textit{R}(\textit{T}) data clearly demonstrates a superconducting transition, as evidenced by a sharp drop at 58 K, reaching zero resistance at around 53 K. For s2, we measured \textit{R}(\textit{T}) under different pressures ranging from 152 to 118 GPa, Fig. 3(b). The $T_\mathrm{c}$ decreases upon decompression, reaching about 51 K at 118 GPa for s2. Upon further decompression, the diamond anvils were cracked. The pressure dependence of $T_\mathrm{c}$ obtained from resistance measurements on these samples is depicted in Fig. 3(c). The highest $T_\mathrm{c}$ observed in this experiment is 62 K, recorded in s3, s4 and s5 at 163 GPa, 164 GPa and 155 GPa, respectively.

\begin{figure}[!t]
	\begin{center}
		\epsfxsize=8.5cm
		\epsffile{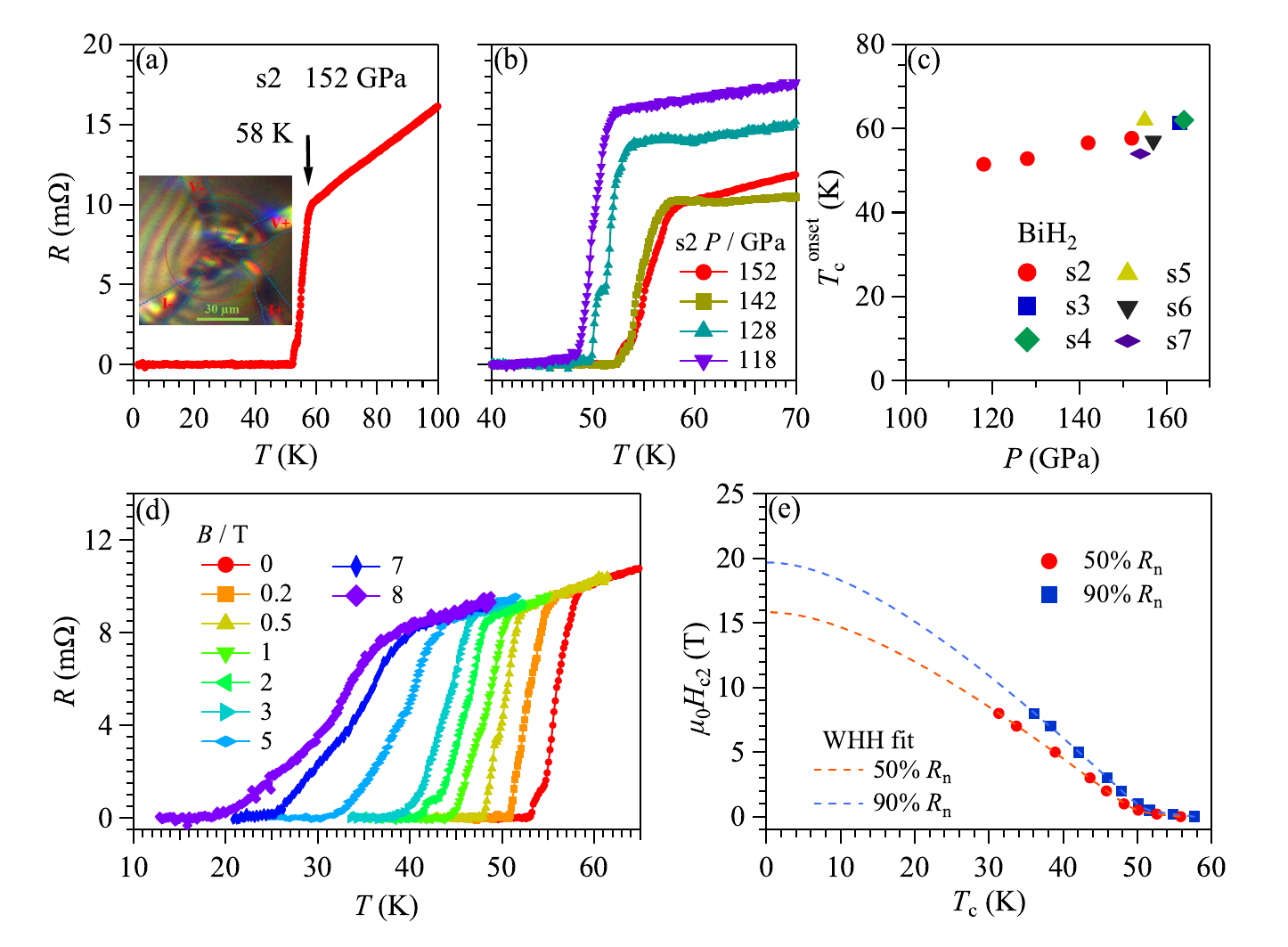}
	\end{center}
	\caption{(Color online) (a) Temperature-dependent resistance of s2 at 152 GPa. The left inset shows the configuration of sample in chamber and the permutation of electrodes. (b) Temperature-dependent resistance of s2 under selected pressures. (c) Pressure dependent $T_\mathrm{c}$ of \textit{Cmcm}-BiH$_{2}$ in different runs. (d) Temperature dependences of resistance under various magnetic fields for s2 at 152 GPa. (e) Temperature dependence of \textit{$\mu$}$_{0}$\textit{H}$_\mathrm{c2}$(T) fitted by the WHH two-band model. }
	\label{fig3}
\end{figure}

 To further substantiate the occurrence of the superconducting transition, we examined the sample's response to external magnetic fields. Figure 3(d) shows the \textit{R}(\textit{T}) of s2 under various magnetic fields. As can be seen, the superconducting transition is progressively suppressed with increasing magnetic fields, in line with the expected downward shift behavior of a superconductor. The values of $T_\mathrm{c}$ at 50\%-\textit{R}$_{n}$ and 90\%-\textit{R}$_{n}$ were utilized to establish the relationship between \textit{$\mu$}$_{0}$\textit{H}$_\mathrm{c2}$(\textit{T}) and temperature. As shown in Fig. 3(e), within the magnetic field range of 0-8 T, \textit{$\mu$}$_{0}$\textit{H}$_\mathrm{c2}$(\textit{T}) exhibits a distinct nonlinear behavior, characterized by a notable upward curvature, a feature commonly observed in two-band superconductors. By fitting the experimental \textit{$\mu$}$_{0}$\textit{H}$_\mathrm{c2}$(\textit{T}) data using the Werthamer-Helfand-Hohenberg (WHH) two-band model in the dirty limit \cite{A23}, we determined the zero-temperature \textit{$\mu$}$_{0}$\textit{H}$_\mathrm{c2}$(0) to be 15.8 T for $T_\mathrm{c}$$^{50\%}$\textit{$^{R}$}$^\mathrm{n}$ and 19.7 T for $T_\mathrm{c}$$^{9}$$^{0\%}$\textit{$^{R}$}$^\mathrm{n}$, respectively (Table S6 \cite{SI}). These values are much smaller than the weak-coupling Pauli limit of \textit{$\mu$}$_{0}$\textit{H}$_\mathrm{p}$$^\mathrm{BCS}$ = 1.84$T_\mathrm{c}$ $\mathrm{\approx}$ 102.7 and 106.2 T. Details of the fitting process are provided in the SM \cite{SI}. Additionally, after detecting superconductivity in sample s5, we conducted in-situ XRD measurements to verify the phase identity. A comparison of the XRD pattern of s5 at 160 GPa with that of s1 at 163 GPa, Fig. S5 \cite{SI}, further confirmed that the superconducting sample is indeed the \textit{Cmcm}-BiH$_{2}$.

The observed $T_\mathrm{c}$ of 62 K at 163 GPa for \textit{Cmcm}-BiH$_{2}$ marks the first instance of superconductor among the \textit{M}H$_{2}$‑type metal dihydrides. To gain deeper insight into the superconducting properties of \textit{Cmcm}-BiH$_{2}$, we further analyzed its phonon DOS, Eliashberg phonon spectral function \textit{$\upalpha$}$^{2}$\textit{F}(\textit{$\omega$}), and EPC constant \textit{$\lambda$}(\textit{$\omega$}). As shown in Fig. 4(a), the absence of imaginary frequencies in the phonon spectra of \textit{Cmcm}-BiH$_{2}$ at 150 GPa indicates its dynamical stability. The calculated phonon DOS can be divided into two regions: low-frequency vibrations from Bi atoms and mid- to high-frequency vibrations from hydrogen atoms. The calculated $\lambda$ and $\omega_\mathrm{log}$ for \textit{Cmcm}-BiH$_{2}$ is around 1.27 and 37.19 meV at 150 GPa. The calculated $T_\mathrm{c}$ of \textit{Cmcm}-BiH$_{2}$ by using the Migdal-Eliashberg equations with \textit{$\mu$}$^{*}$ = 0.1 is about 54 K at 150 GPa, which is close to our experimental observation.

\begin{figure}[!t]
	\begin{center}
		\epsfxsize=8.5cm
		\epsffile{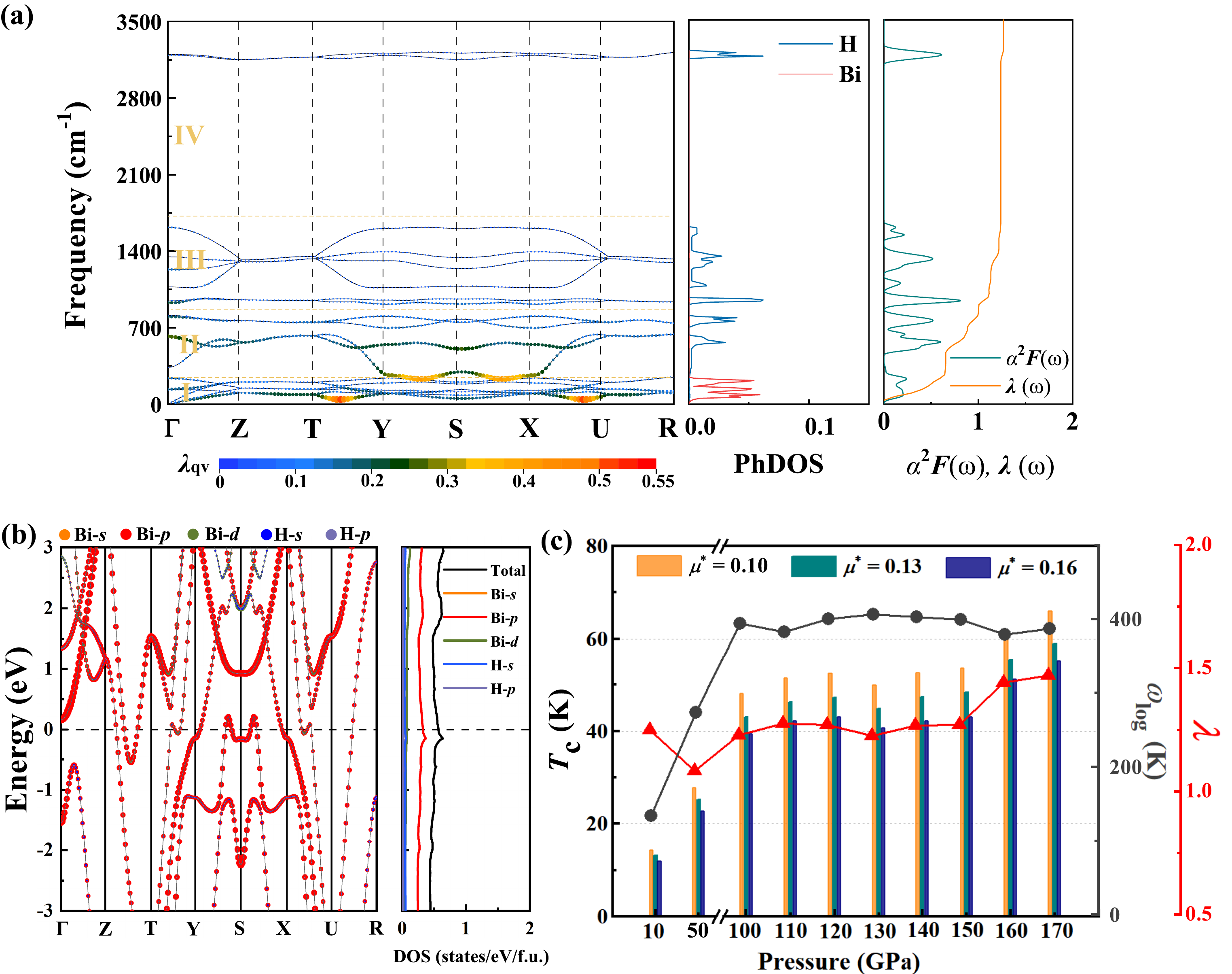}
	\end{center}
	\caption{(Color online)   (a) Calculated phonon dispersion, projected phonon density of states (PhDOS), Eliashberg phonon spectral function $\upalpha$$^{2}$\textit{F}(\textit{$\omega$}), and integrated electron-phonon coupling \textit{$\lambda$}(\textit{$\omega$}) for \textit{Cmcm}-BiH$_{2}$ at 150 GPa. (b) The calculated electronic band structure and projected DOS of \textit{Cmcm}-BiH$_{2}$ at 150 GPa. (c) The calculated electron-phonon coupling constant $\lambda$, logarithmic average frequency $\omega_\mathrm{log}$ and $T_\mathrm{c}$ for \textit{$\mu$}$^{*}$ = 0.1, 0.13 and 0.16 at the pressure range of 10 to 170 GPa. }
	\label{fig4}
\end{figure}

As illustrated in Fig. 4(a), the Bi-derived and H-derived phonon modes are well separated. Thus, the total EPC constant $\lambda$ can be decomposed into atom-specific contributions, $\lambda_\mathrm{Bi}$ and $\lambda_\mathrm{H}$. Specifically, the phonon dispersion curves are divided into four regions according to the vibrational frequencies, i.e., \uppercase\expandafter{\romannumeral1}, \uppercase\expandafter{\romannumeral2}, \uppercase\expandafter{\romannumeral3}, and \uppercase\expandafter{\romannumeral4} in Fig. 4(a), and the detailed vibrational modes of each region are demonstrated in Fig. S6 of the SM \cite{SI}. The region \uppercase\expandafter{\romannumeral1} in the vibrational frequencies from 0 to 250 cm$^{-}$$^{1}$ is primarily attributed to Bi atoms, yielding an integrated $\lambda_\mathrm{Bi}$ $\sim$ 0.65, which contributes approximately 51\% of the total $\lambda$ in \textit{Cmcm}-BiH$_{2}$. In this region, the phonon modes are predominantly associated with the movements of Bi atoms, including in-phase vibrations at 0 cm$^{-}$$^{1}$ (acoustic branch) and bending vibrations within the Bi lattice at 122-250 cm$^{-}$$^{1}$ (relative vibrations between pairs of Bi atoms). Regions \uppercase\expandafter{\romannumeral2} and \uppercase\expandafter{\romannumeral3} in the frequencies ranging from 250 to 1640 cm$^{-}$$^{1}$, contribute approximately 46\% to EPC. These regions involve the vibrations of H ions perpendicular to Bi-H bonds and the in-phase vibrations of H$_{2}$ molecules, respectively. In Region \uppercase\expandafter{\romannumeral4}, the vibrations of the hydrogen atoms along the H-H bond direction contribute only 3\% to EPC. As shown in Fig. 4(b), the calculated electronic structure exhibits metallic behavior with a relatively high \textit{N}($E_\mathrm{F})$. The phonon dispersion demonstrates that the strongest EPC contributions arise along the T-Y and X-U paths, with additional significant contributions from the Y-S-X path. Notably, the electronic bands near the Fermi level along these paths are predominantly derived from ${p}$ orbitals of Bi. Consistently, the projected DOS analysis shows that ${p}$ states of Bi account for the majority ($\sim$ 79\%) of \textit{N}($E_\mathrm{F})$, ensuring strong coupling with low-frequency phonon modes involving Bi vibrations. However, the contribution from hydrogen-derived states is relatively small. This is consistent with the behavior typically observed in molecular hydrides, where the electronic states near the Fermi level are dominated by the heavy element sublattice.

According to the McMillan-Hopfield theory \cite{A25, A26}, the large $\lambda_\mathrm{Bi}$ mainly originates from the high \textit{N}($E_\mathrm{F})$(Bi) contributions, despite the smaller \textit{I}$^{2}$ and larger \textit{M}$\omega_\mathrm{2}$$^{2}$ values compared to H (Table S5 \cite{SI}). In contrast, in covalent and clathrate-type hydride systems, both the \textit{N}($E_\mathrm{F})$ and the EPC strength are typically dominated by hydrogen-derived electronic and vibrational contributions \cite{A9, A10, A24}. Although hydrogen makes a relatively small contribution to \textit{N}($E_\mathrm{F})$ in \textit{Cmcm}-BiH$_{2}$, the presence of H$_{2}$ molecules plays a crucial role in boosting $\omega_\mathrm{log}$ while maintaining the strong $\lambda$, thereby further enhancing superconductivity of BiH$_{2}$. This essential role of incorporating of H$_{2}$ molecules explains why the $T_\mathrm{c}$ of \textit{Cmcm}-BiH$_{2}$ ($\sim$ 62 K) is significantly higher compared to Bi-III phase ($\sim$ 7 K) \cite{A13}, even though the total $\lambda$ of \textit{Cmcm}-BiH$_{2}$ is lower than that of Bi-III ( 2.75). Notably, the $\omega_\mathrm{log}$ of \textit{Cmcm}-BiH$_{2}$ is around ten times higher than that of Bi-III (3.73 meV). Such synergy leads to an increase in $\omega_\mathrm{log}$, ultimately resulting in a substantial elevation of $T_\mathrm{c}$. Remarkably, \textit{Cmcm}-BiH$_{2}$ remains dynamically stable down to 10 GPa without imaginary frequencies on the phonon spectrum (Fig. S7 \cite{SI} ), a significantly lower pressure than typically required to stabilize high-pressure hydride superconductors. As pressure decreases, $T_\mathrm{c}$ slowly and monotonically decreases, remaining around 48 K at 100 GPa. However, below 100 GPa, the $T_\mathrm{c}$ decreases rapidly, reaching approximately 17 K at 10 GPa [Fig. 4(c)].

It should be noted that the decomposition of BH$_{3}$NH$_{3}$ is a complex multistep process and its direct contact with Bi during high-pressure/high-temperature (HPHT) synthesis might potentially produce other phases containing B and/or N. To address this issue, we conducted comprehensive control experiments as follows. Firstly, we test directly the possible reaction between Bi and BN by treating their mixtures under comparable HPHT conditions as those for the BiH$_{2}$ synthesis. Synchrotron XRD measurements (Fig. S8 \cite{SI}) revealed no formation of borides, nitrides, or related phases, confirming that Bi and BN do not react under these conditions. Secondly, we carried out HPHT syntheses by using alternative hydrogen sources free of boron and nitrogen, i.e., paraffin and pure molecular hydrogen. Synchrotron XRD confirmed that the same crystalline phase was consistently obtained using these alternative hydrogen sources, identical to that synthesized with BH$_{3}$NH$_{3}$ (Figs. S8 and S9 \cite{SI}). Furthermore, electrical transport measurements on two paraffin-derived samples revealed superconducting transitions at ~57 K and ~54 K, respectively (Fig. S8 \cite{SI}), in excellent agreement with the pressure-dependent $T_\mathrm{c}$ values observed in BH$_{3}$NH$_{3}$-derived samples (Fig. 3c). Taken together, these two sets of experiments conclusively rule out any chemical incorporation or reaction involving B or N with the superconducting phase, providing unambiguous confirmation that the synthesized superconducting compound is indeed \textit{Cmcm}-BiH$_{2}$.

According to recent theoretical study \cite{A28}, the $T_\mathrm{c}$ of BiH$_{2}$ will not exceed 58 K, due to the relatively high atomic mass ratio between Bi and H. Thus, the experimental observation of a high $T_\mathrm{c}$ up to 62 K in \textit{Cmcm}-BiH$_{2}$ is quite unexpected for a heavy metal dihydride with such a low hydrogen content. It is well known that \textit{M}H$_{2}$-type dihydride is one of the most common forms of metal hydrides at ambient pressure, typically characterized as insulators or poor metals. In these conventional \textit{M}H$_{2}$ compounds, the structure is often ionic, with hydrogen atoms occupying interstitial octahedral or tetrahedral sites within densely packed metal lattices. This configuration leads to the localization of hydrogen electrons at low energy levels, contributing minimally to \textit{N}($E_\mathrm{F})$. To our knowledge, metal dihydrides \textit{M}H$_{2}$ have never been found to exhibit superconductivity. Unlike previously reported high-$T_\mathrm{c}$ hydride superconductors \cite{A9, A10, A24}, \textit{Cmcm}-BiH$_{2}$ represents a novel structural type where a three-dimensional bismuth host framework with weak covalent Bi-Bi interactions encapsulates H$_{2}$-like molecules as guests. DFT calculations further reveal that the unique bismuth open-channel framework forms metallic conduction channels, dominating the electronic states near the Fermi level and playing a crucial role in its superconducting properties. These findings underscore the pivotal role of non-hydrogen elements in hydride superconductors and offer fresh insights and guidance for developing high-$T_\mathrm{c}$ hydride superconductors at lower pressures.

In summary, we have successfully synthesized a bismuth dihydride superconductor, \textit{Cmcm}-BiH$_{2}$, featuring a unique host-guest structure. In \textit{Cmcm}-BiH$_{2}$, weak Bi-Bi covalent bonds form a three-dimensional bismuth open-channel framework that encapsulates H$_{2}$-like molecules as guests. \textit{Cmcm}-BiH$_{2}$ exhibits superconductivity with $T_\mathrm{c}$ of 62 K at 163 GPa, establishing the first reported instance of superconductor among \textit{M}H$_{2}$‑type metal dihydrides. Detailed analyses reveal that the bismuth open-channel framework play a dominant role in its superconducting properties. Our work not only uncovers a novel hydride structure, broadening the scope of experimentally realized metal hydrides, but also provides valuable insights for the design and optimization of new hydride superconductors.

${}$

\textit{Acknowledgments}—This work is supported by the National Key Research and Development Program of China (2021YFA1400200, 2023YFA1406100, 2023YFA1608902, 2022YFA1402301), National Natural Science Foundation of China (12025408, 12304030, 12347157, 11888101, 12247177, U23A6003), CAS Project for Young Scientists in Basic Research (2022YSBR-047), and CAS PIFI program (2024PG003), the Project of China Postdoctoral Science Foundation under Grant Number (2023M743223, 2025T180925), the Postdoctoral Fellowship Program of CPSF under Grant Number (GZB20230674, GZC20252191), and Postdoctoral Innovation Talents Support Program of Henan (2023032). We thank the staff members of the BL15U1 station (https://cstr.cn/31124.02.SSRF.BL15U1) and the Shanghai Synchrotron Radiation Facility of Experiment Assist System (https://cstr.cn/31124.02.SSRF.LAB), 4W2 station (https://cstr.cn/31109.02.BSRF.4W2) in Beijing Synchrotron Radiation Facility (BSRF) (https://cstr.cn/31109.02.BSRF), BL10XU station in Super Photon ring-8 (Spring-8), for providing technical support and assistance in data collection and analysis. Part of the high-pressure resistance measurements were performed at the Cubic Anvil Cell station (https://cstr.cn/31123.02.SECUF.A2) of Synergic Extreme Condition User Facility (SECUF) (https://cstr.cn/31123.02.SECUF).

${}$
${}$

\textit{Data availability}—The data that support the findings of this article are not publicly available. The data are available from the authors upon reasonable request.

${}$

Liang Ma, Xin Yang and Mei Li equally contributed to this work.

$^{\ast}$ Electronic address: pengfei.shan@hpstar.ac.cn

$^{\dagger}$ Electronic address: hguo@zzu.edu.cn

$^{\ddagger}$ Electronic address: jgcheng@iphy.ac.cn


\end{document}